\documentclass[aps,prb,showpacs,tightenlines,twocolumn,nofootinbib,nobibnotes,superscriptaddress]{revtex4-1}
\usepackage{amsmath,amssymb,amsfonts,bm}
\usepackage{graphicx}
\usepackage{epstopdf}
\usepackage{dcolumn}
\usepackage{mathrsfs}
\usepackage{bbold}
\usepackage{dsfont}
\usepackage{ulem}
\usepackage[colorlinks=true,linkcolor=blue,citecolor=blue, urlcolor=blue,bookmarks=false]{hyperref}

\begin{document}

\title{Finite-size effects in non-Hermitian topological systems}
\author{Rui Chen}
\affiliation{Department of Physics, Hubei University, Wuhan 430062, China}
\author{Chui-Zhen Chen}
\affiliation{Institute for Advanced Study and School of Physical Science and Technology, Soochow University, Suzhou 215006, China}
\affiliation{Department of Physics, Hong Kong University of Science and Technology, Clear Water Bay, Hong Kong, China}
\author{Bin Zhou}\thanks{binzhou@hubu.edu.cn}
\author{Dong-Hui Xu} \thanks{donghuixu@hubu.edu.cn}
\affiliation{Department of Physics, Hubei University, Wuhan 430062, China}
\begin{abstract}

We systematically investigate the finite-size effects in non-Hermitian one-dimensional (1D) Su-Schrieffer-Heeger (SSH) and two-dimensional (2D) Chern insulator models. In the Hermitian SSH system, the finite-size energy gap is always real and shows a monotonic-exponential decay as the chain length grows. In contrast, for the non-Hermitian SSH model, the non-Hermitian intra-cell hoppings can modify the localization lengths of bulk and end states, giving rise to a complex finite-size energy gap that exhibits an oscillating exponential decay as the chain length grows. However, the imaginary staggered on-site potentials in the SSH model only change the end-state energy, leaving the localization lengths of the system unchanged. In this case, the finite-size energy gap can undergo a transition from real values to imaginary values. We observed similar phenomena for the finite-size effect in 2D non-Hermitian Chern insulator systems.

\end{abstract}

\maketitle

\section{Introduction}

In conventional quantum mechanics, physical observations are represented by Hermitian operators whose eigenvalues are always real. Nevertheless, in open quantum systems that exchange energy and/or matter with their environments, non-Hermitian operators are proved to be particularly useful~\cite{Dittes2000PR,Heiss2012JPAMT,Bender1998PRL,Bender2007RPP}. The Hamiltonian operators of these systems are non-Hermitian. Their energy eigenvalues could be complex, and the imaginary part is also related to experimentally observable quantities~\cite{Feshbach1954PR,ElGanainy2018NatPhys}. Over the past decades, remarkable theoretical and experimental progresses have been achieved in various non-Hermitian systems, such as open quantum systems~\cite{Lee2014PRX,Cao2015RMP,Rotter2009JPAMT,Carmichael1993PRL}, optical systems with gain and loss~\cite{Feng2014Science,Hodaei2017nature,Regensburger2012Nature,Peng2014Science,Cerjan2016PRA,Guo2009PRL}, and interacting\cite{SINHA2005MPLA,CastroAlvaredo2009JPA} or disordered\cite{Zeng2017PRA,Heinrichs2001PRB,Goldsheid1998PRL,Hatano1996PRL,Hatano1997PRB} systems.

At the mean time, the study of topological phases of quantum matter~\cite{Bernevig2013TI,Qi2011RMP,Chiu2016RMP,Shen2017TI}, such as topological insulators~\cite{Bansil2016RMP,Liu2016AnnualReview,Hasan2010RMP,Ando2013JPSJ}, topological superconductors~\cite{Alicea2012RPP,Stanescu2013JPCM,Elliott2015RMP}, and topological semimetals~\cite{Armitage2018RMP}, has become one of the major fields in condensed matter physics. A hallmark of these topological phases is the symmetry-protected topological gapless boundary states, which have attracted enormous attention because of their exotic properties and potential applications in electronic devices.

Very recently, the aforementioned two fields---non-Hermitian physics and symmetry-protected topological phases---start to merge together~\cite{Bandres2018Phys,alvarez2018topological,Rudner2009PRL}. Dissipative topological transitions~\cite{Zeuner2015PRL} and topologically protected boundary states~\cite{Weimann2016NatMater,Harari2018Science,Bandres2018Science} have been confirmed experimentally in the non-Hermitian optical systems. Meanwhile, non-Hermitian physics has been intensively investigated in other topological systems, such as topological insulators~\cite{Yao2018PRLa,Yao2018PRLb,Esaki2011PRB,Lee2016PRL,Lieu2018PRB}, topological semimetals ~\cite{Xu2017PRL,Cerjan2018PRB,Cerjan2018arXiv,Zyuzin2018PRB, Wang2018arXiv,Yang2018Arxiv,Lee2018arXiv2} and topological superconductors~\cite{SanJose2016SciRep,Kawabata2018PRB1, Avila2018arXiv,Zyuzin2019arXiv}. In non-Hermitian systems, the interplay between non-Hermiticity and topology produces interesting phenomena that have no counterparts in Hermitian systems, such as the breakdown of the bulk-boundary correspondence~\cite{Leykam2017PRL,Gong2018PRX,Shen2018PRL,Jin2018arXiv,Xiong2018JPC,Hu2011PRB,Kunst2018PRL}, the emergence of anomalous edge states~\cite{,Lee2016PRL,Kawabata2018PRB}, the non-Hermitian skin effect~\cite{Yao2018PRLa,MartinezAlvarez2018PRB,Lee2018arXiv1,Lee2018arXiv2}, as well as the deviation of the Hall conductance from its quantized value in non-Hermitian Chern insulators\cite{Philip2018PRB,Chen2018arXiv}.
Currently, most of the studies are concentrated on infinite or semi-infinite systems. However, it is essential to study the case of a finite chain or strip geometry which is used in experiments.

In this work, we study finite-size effects in non-Hermitian topologically gapped systems. The finite-size effects had been well theoretically and experimentally studied in Hermitian topological systems, such as topological insulators \cite{Zhou2008PRL,Chen2016CPB,JiangHua2014,Linder2009PRBa,Zhang2010NatPhys,LuHZ2010PRB,LiuCX2010PRB,Imura2012PRB} and   semimetals~\cite{Takane2016JPSJ,Chen2017PRBDirac,Wang2012PRB,Wang2013PRB,Xiao2015SciRep,
Pan2015SciRep,Schumann2018PRL,Yilmaz2017arXiv,Collins2018Nature}. A common feature in these systems is that the energy gap opened by the hybridization of gapless boundary states instead of symmetry breaking could exhibit a monotonic or an oscillating exponential decay with increasing the system size. 
In order to monitor how non-Hermiticity affects the finite-size effects in topological systems, by combining numerical and analytical methods, we investigate the finite-size effect in one-dimensional (1D) Su-Schrieffer-Heeger (SSH) and two-dimensional (2D) Chern insulator models subjected to various non-Hermitian terms.

For a finite Hermitian SSH chain with open boundaries, the finite-size energy gap displays a purely exponential decay as the system size increases. We found that non-Hermiticity has significant influences on the finite-size effect in the SSH model. The  non-Hermitian intra-cell hopping terms that respect chiral symmetry can modify the localization lengths of the end and/or bulk states, giving rise to a complex finite-size energy gap which exhibits an oscillating exponential decay as the chain length grows. However, the non-Hermitian staggered on-site potentials that break chiral symmetry only change the end-state energy, leaving the localization lengths of the system unchanged. In this case, the finite-size energy gap could undergo a transition from real values to imaginary values. We also study the finite-size effect in a 2D Chern insulator model and found similar phenomena as in the 1D SSH model.

This paper is organized as follows: In Sec.~\ref{SSH-Model},
we introduce a model Hamiltonian describing the non-Hermitian SSH model and perform an analytical study on the finite-size effect. Then, we present both the numerical and analytical results of the finite-size effect in Sec.~\ref{SSH-Results}. In Sec.~\ref{QWZ}, we mainly study the finite-size effect in a 2D non-Hermitian Chern insulator model. Finally, a brief summary is presented in Sec.~\ref{Conclusion}.

\section{Non-Hermitian SSH model}
\label{SSH-Model}

\begin{figure}[ptb]
\includegraphics[width=8cm]{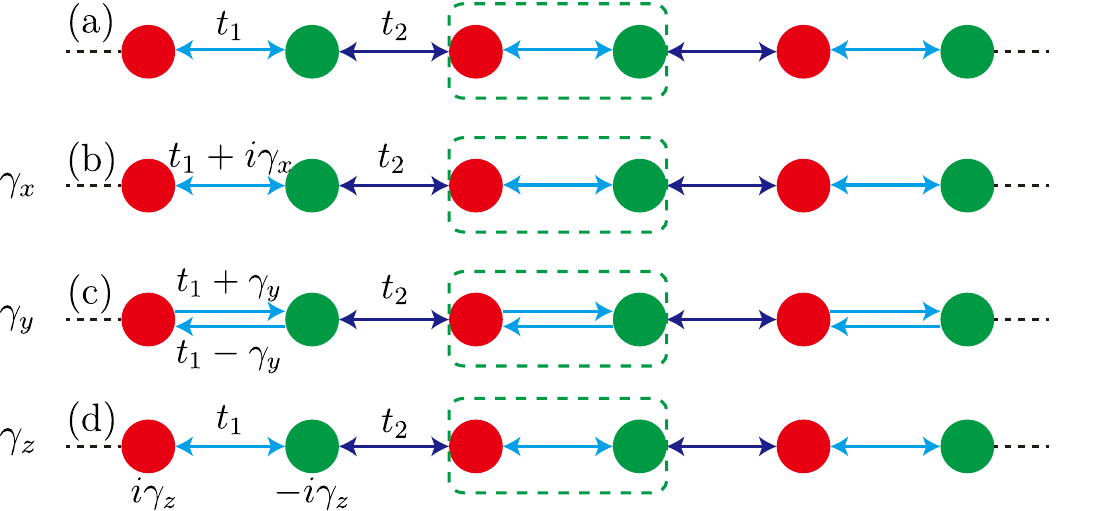}
\caption{(Color online) Schematic illustrations of the (a) Hermitian and (b-d) non-Hermitian SSH model chains. The dashed box encloses a unit cell that contains two sublattices marked by red and green filled circles. The blue arrowed lines identify the inter-cell hoppings, and the hopping strengths are fixed as $t_2=1$. The cyan arrowed lines correspond to the intra-cell hoppings, and the hopping strengths are (a) $t_1$, (b) $t_1+ i\gamma_x$, (c) $t_1\pm\gamma_y$, and (d) $t_1$, respectively. Here, the non-Hermiticity is manifested by (b) non-conjugated intra-cell hoppings, (c) unequal intra-cell hoppings, and (d) imaginary staggered on-site potentials $\pm i \gamma_z$. }
\label{fig1}
\end{figure}

We start with a non-Hermitian SSH model defined on 1D dimer chain consisting of two sublattices as depicted in Fig.~\ref{fig1}. The realization of non-Hermitian SSH models has been reported in  a series of experiments recently~\cite{Meier2018Science,Weimann2016NatMater,Zeuner2015PRL,ElGanainy2018NatPhys,Poli2015NatCom}. The Bloch Hamiltonian of the non-Hermitian SSH model is given by
\begin{equation}
H=H_{\text{SSH}}+H_{\gamma},
\label{SSH-model}
\end{equation}
where $H_{\text{SSH}}\left( k\right) =d_{x}\sigma _{x}+d_{y}\sigma _{y}$ is the conventional Hermitian SSH model [Fig.~\ref{fig1}(a)] with $d_{x}=t_{1}+t_{2}\cos k$, $d_{y}=t_{2}\sin k$. $\sigma _{i}$'s are the Pauli matrices acting on the sublattice subspace. $t_1$ and $t_2$ describe the intra-cell and inter-cell hopping strengths, respectively. We fix $t_2=1$ in the following calculations unless otherwise noted. $H_{\gamma}=i\left( \gamma _{x}\sigma _{x}+\gamma _{y}\sigma _{y}+\gamma
_{z}\sigma _{z}\right)$ depicts the non-Hermitian terms with $\gamma _{x}$ and $\gamma _{y}$ describing the non-conjugated and unequal intra-cell hoppings [Figs.~\ref{fig1}(b-c)] and $\gamma _{z}$ describing the imaginary staggered on-site potentials [Fig.~\ref{fig1}(d)]. The eigenvalues of the Hamiltonian (\ref{SSH-model}) can be written as $E_{\pm }\left( k\right) =\pm \sqrt{%
\left( d_{x}+i\gamma _{x}\right) ^{2}+\left( d_{y}+i\gamma _{y}\right)
^{2}-\gamma _{z}^{2}}$.

This non-Hermtian SSH model still respects chiral symmetry $\sigma_{z}H\left(k\right)\sigma_{z}=-H\left(k\right)$ as long as $\gamma_z=0$. Much efforts have been devoted to developing proper topological invariants to restore the bulk-boundary correspondence in the non-Hermitian SSH model, such as the $\mathbb{Z}$ invariant in terms of the complex Berry phase~\cite{Lieu2018PRB}, the biorthogonal polarization $P$~\cite{Kunst2018PRL}, the non-Bloch winding number $W$~\cite{Yao2018PRLa} and the singular-value description~\cite{Herviou2019arXiv}. When $\gamma_{x,y}=0$, $\gamma_z\neq 0$, chiral symmetry in the SSH model is broken, but it possesses $PT$ symmetry with $\sigma_x H\left(k\right) \sigma_x=H\left(k\right)^*$~\cite{Kunst2018PRL,Lieu2018PRB}. Actually, the $PT$ symmetric non-Hermtian SSH model has been realized in recent experiments based on coupled optical waveguides~\cite{Zeuner2015PRL,Poli2015NatCom,Weimann2016NatMater,ElGanainy2018NatPhys,Zhao2018NatCon}, and its topological properties have also been studied by researchers~\cite{Schomerus2013OL,Lieu2018PRB,Kunst2018PRL}. Moreover, in the parameter space, $PT$ symmetric models have two regions referred to as the unbroken and broken phases: the former one is characterized by a fully real spectrum, and the eigenfunctions of the Hamiltonian are also eigenfunctions of the $PT$ operator, while the latter one possesses complex energies and the corresponding eigenfunctions of the Hamiltonian are not eigenfunctions of the $PT$ operator.~\cite{Bender1998PRL,Bender2007RPP}

\subsection{Semi-infinite chains }
\label{SSH-left}
Considering a semi-infinite non-Hermitian SSH chain with a left boundary, the Schr\"{o}dinger equation is given by
\begin{equation}
\left( H^{\text{L}}-E\right) \Psi ^{\text{L}}=%
\begin{pmatrix}
\epsilon & T &  &  &  \\
T^{\dag } & \epsilon & T &  &  \\
& \ddots & \ddots & \ddots &  \\
&  & T^{\dag } & \epsilon & T \\
&  &  & \ddots & \ddots%
\end{pmatrix}%
\begin{pmatrix}
\Psi _{1}^{\text{L}} \\
\Psi _{2}^{\text{L}} \\
\vdots \\
\Psi _{j}^{\text{L}} \\
\vdots%
\end{pmatrix}%
=0,
\label{Seqns}
\end{equation}
where $H^{\text{L}}$ is the corresponding tight-binding Hamiltonian, $\epsilon =%
\begin{pmatrix}
-E_{-} & t_{+} \\
t_{-} & -E_{+}
\end{pmatrix}$
, $T=%
\begin{pmatrix}
0 & 0 \\
t_{2} & 0%
\end{pmatrix}$
, $E_{\pm}=E\pm i\gamma_z$, $t_{\pm }=t_{1}+i\gamma _{x}\pm \gamma _{y}$, $\Psi _{i}^{\text{L}}=%
\begin{pmatrix}
\Psi _{A,i}^{\text{L}} \\
\Psi _{B,i}^{\text{L}}%
\end{pmatrix}%
$, and $\Psi _{A/B,i}$ are the wave function amplitudes on sublattices A and B of the $i$th lattice site.

To solve the  Schr\"{o}dinger equation, we take the ansatz $\Psi _{i}^{\text{L}}=\rho_{\text{L}}  \Psi _{i-1}^{\text{L}}=\rho _{\text{L}}^{i}\Psi _{0}^{\text{L}}$\cite{Chen2016CPB,Creutz1994PRD,Creutz2001RMP,Konig2008JPSP,Yao2018PRLa}, where the quantity $\rho _{\text{L}}$ is a complex number. With this, Eq. (\ref{Seqns}) can be reduced to the following two equations:
\begin{eqnarray}
\rho _{\text{L}}\left( \epsilon +\rho _{\text{L}}T\right) \Psi _{0}^{\text{L}} &=&0 , \nonumber \\
\rho _{\text{L}}\left( T^{\dag }+\rho _{\text{L}}\epsilon +\rho _{\text{L}}^{2}T\right) \Psi
_{0}^{\text{L}} &=&0.
\end{eqnarray}%
By solving the two equations, we can
obtain the energy $E_{\text{L}}^{\text{e}}$, $\rho _{\text{L}}^{\text{e}}$ and $\Psi _{0}^{\text{e},\text{L}}$ for the end state at the left boundary
\begin{equation}
E_{\text{L}}^{\text{e}}=i\gamma _{z},\rho _{\text{L}}^{\text{e}}=-\frac{t_{-}}{t_{2}},\Psi _{0}^{\text{e},\text{L}}=\begin{pmatrix}
1 \\
0%
\end{pmatrix}.%
\label{Left}
\end{equation}%
We can see that $\Psi^{\text{e},\text{L}}_{B,0}$ component is vanishing for the end state. Note that the analytical expression of $\rho _{\text{L}}^{\text{e}}$ for the end state is in accordance with the result obtained by using a finite SSH chain terminated with the same type of sublattice at both two ends \cite{Kunst2018PRL}.

In addition, we can obtain $\rho _{\text{L}}^{\text{b}}$ for the bulk states
\begin{equation}
\rho _{\text{L}}^{\text{b}}=\pm \sqrt{\frac{t_{-}}{t_{+}}}.
\end{equation}
The detailed formulas of energies and wave functions of bulk states are omitted since they are complicated and not important in this work.

In a similar way we can solve the Schr\"{o}dinger equation for a semi-infinite non-Hermitian SSH chain with the endpoint located at the right side. The energy $E_{\text{R}}^{\text{e}}$, $\rho _{\text{R}}^{\text{e},\text{b}}$, and $\Psi _{0}^{\text{e},\text{R}}$ for the end state are found as follows
\begin{align}
E_{\text{R}}^{\text{e}}=-i\gamma _{z},\rho _{\text{R}}^{\text{e}}&=-\frac{t_{+}}{t_{2}},\Psi _{0}^{\text{e},\text{R}}=%
\begin{pmatrix}
0 \\
1%
\end{pmatrix},
\notag
\\
\rho _{\text{R}}^{\text{b}}&=\pm \sqrt{\frac{t_{+}}{t_{-}}}.\label{Right}
\end{align}%

From Eqs. (\ref{Left}) and (\ref{Right}), we can clearly see that the non-Hermitian term $\gamma_z$ only modifies the end-state energy $E_{\text{L},\text{R}}^{\text{e}}$, while $\gamma_x$ and $\gamma_y$ affect $\rho _{\text{L},\text{R}}^{\text{e},\text{b}}$. Since the absolute values of $\rho _{\text{L},\text{R}}^{\text{e},\text{b}}$ are related to the localization lengths of the end and bulk states, adding $\gamma_x$ and $\gamma_y$ will adjust the localization lengths of the system. The smaller $\left|\rho _{\text{L},\text{R}}^{\text{e}}\right|$ and $\left|\rho _{\text{L},\text{R}}^{\text{b}}\right|$ are, the more localized the wave functions of the edge and bulk states become. In contrast with the Hermitian case, the localization lengths for the end states at the left and right boundaries are not equivalent anymore in the presence of $\gamma_{y}$.

It is worthy of noting that the appearance of the topological end states
requires $\left\vert \rho _{\text{L}\left( \text{R}\right) }^{\text{e}}\right\vert
<\left\vert \rho _{\text{L}\left( \text{R}\right) }^{\text{b}}\right\vert $~\cite{Yao2018PRLa}, otherwise, the end states merge into the bulk states. For
$\gamma_x\neq 0$ and $\gamma_y=\gamma_z=0$, the end state would appear when $\left\vert
t_{1}\right\vert <\sqrt{t_{2}^{2}-\gamma _{x}^{2}}$ is satisfied. For $\gamma_y\neq 0$ and $\gamma_x=\gamma_z=0$, the end state appears when $\left\vert
t_{1}\right\vert <\sqrt{t_{2}^{2}+\gamma _{y}^{2}}$ if $\left\vert
t_{2}\right\vert >\left\vert \gamma _{y}\right\vert $, and $\sqrt{%
-t_{2}^{2}+\gamma _{y}^{2}}<\left\vert t_{1}\right\vert <\sqrt{%
t_{2}^{2}+\gamma _{y}^{2}}$ if $\left\vert t_{2}\right\vert <\left\vert
\gamma _{y}\right\vert $. In the case of $\gamma_z\neq 0$ and $\gamma_x=\gamma_y=0$, comparing with the Hermitian
case, the localization lengths remain unchanged and the end state appears when $\left\vert t_{1}\right\vert <\left\vert
t_{2}\right\vert $.

For a generic non-Hermitian Hamiltonian $H$, the eigenvalue equations have the following forms:
\begin{equation}
H\left\vert \Psi _{\mathcal{R}}\right\rangle =E\left\vert \Psi _{\mathcal{R}}\right\rangle
,H^{\dag }\left\vert \Psi _{\mathcal{L}}\right\rangle =E^{\ast }\left\vert \Psi
_{\mathcal{L}}\right\rangle,
\end{equation}
where $\left\vert \Psi _{\mathcal{L},\mathcal{R}}\right\rangle$ are the left and right eigenvectors, and they are not simply related by complex-conjugate transpose. Through this paper, we focus on the results obtained from the right eigenvectors. While for the case of the left eigenvectors, the results would be different, for example, the analytical expression of $\rho^{\text{e}}$ for the end states \cite{Kunst2018PRL}. However, in this work, we mainly concentrate on the finite-size effect in the energy spectra of topological non-Hermitian systems, and the eigenvalues for left and right eigenvector are complex conjugate to each other. Therefore, the essential physics remains unchanged for the case of left eigenvectors.

\subsection{Finite-size chain}
Let us now consider a finite non-Hermitian SSH chain with length $n$, and then the Schr\"{o}dinger equation becomes
\begin{equation}
\left( H^{\text{F}}_n-E\right) \Psi^n =%
\begin{pmatrix}
\epsilon & T &  &  &  \\
T^{\dag } & \epsilon & T &  &  \\
& \ddots & \ddots & \ddots &  \\
&  & T^{\dag } & \epsilon & T \\
&  &  & T^{\dag } & \epsilon%
\end{pmatrix}%
\begin{pmatrix}
\Psi _{1}^n \\
\Psi _{2}^n \\
\ddots \\
\Psi _{n-1}^n \\
\Psi _{n}^n%
\end{pmatrix}%
=0,
\label{SSH-chain}
\end{equation}%
where $H^{\text{F}}_n$ is a $2n \times 2n$ square matrix.
We denote the determinant of the secular equation as $F(n)
=\left\vert H^{\text{F}}_n-E\right\vert $, and $F$ satisfies the following recursion relation
\begin{equation}
\left(
\begin{array}
[c]{c}%
F\left(  i\right)  \\
G\left(  i\right)
\end{array}
\right)  =\left(
\begin{array}
[c]{cc}%
\alpha & t_{2}E_{-}  \\
-t_{2}E_{+}   & -t_{2}^{2}%
\end{array}
\right)  \left(
\begin{array}
[c]{c}%
F\left(  i-1\right)  \\
G\left(  i-1\right)
\end{array}
\right),
\end{equation}
where $i=1, 2, \cdots, n$, $\alpha =E_{+}E_{-}-t_{+}t_{-}$, $G\left(i\right)$ is an intermediate variable. Note that the initial values are $%
F\left( 2\right) =\alpha ^{2}-t_{2}^{2}E_{+}E_{-}$, $G\left( 2\right) =-t_{2}E_{+} \left( \alpha
-t_{2}^{2}\right) $. The general expression of $F\left( n\right) $ can be expressed as
$F\left( n\right) =\left(\Lambda _{+-}^{n}\Lambda _{++}-\Lambda
_{--}^{n}\Lambda _{-+}\right)/\left(2^{n+1}\beta \right)$,
where $\Lambda _{\pm \left( \pm \right) }=\alpha \pm \beta \left(\pm\right) t_{2}^{2}$,
and $\beta=\sqrt{\left( \alpha -t_{2}^{2}\right) ^{2}-4t_{2}^{2}t_{+}t_{-}}$. By solving $F\left(n\right)=0$ and expanding at $E=0$, we obtain the desired formula for the end-state energy as follows
\begin{equation}
E^2+\gamma _{z}^{2}=\left(\frac{ t_{2}^{2}-t_{+}t_{-}}{t_{2}}\right)^2
\left(\rho _{\text{L}}^{\text{e}} \rho _{\text{R}}^{\text{e}}\right) ^{n}.
\label{SSH-gap}
\end{equation}
The corresponding wave function can be written as a superposition of the two end states at the left and right boundaries
\begin{equation}
\Psi^{\text{e}}_{i}=C_{1}\Psi _{0}^{\text{e},\text{L}}\left(\rho_{\text{L}}^{\text{e}}\right)^i+C_{2}\Psi _{0}^{\text{e},\text{R}}\left(\rho_{\text{R}}^{\text{e}}\right)^i,%
\label{SSH-wave}
\end{equation}
where $C_{1,2}$ are normalization constants.

It is important to mention that the transfer matrix method we use can also be applied to generic 1D models since there is no particular restriction on this method. In Appendix \ref{Appendix_NNN}, we present the recursion relation of the secular equation for a non-Hermitian SSH model including the next nearest hopping term. However, the explicit solution is not available since the transfer matrix becomes much more complicated.
\section{Results}
\label{SSH-Results}
In this section, we investigate the finite-size effect in the non-Hermitian SSH system by combining numerical calculations and analytical results obtained in the previous section. Before continuing to study the finite-size effect in the non-Hermitian SSH model, we first give the results of the finite-size effect for the Hermitian case.

\begin{figure}[tbp]
\includegraphics[width=8cm]{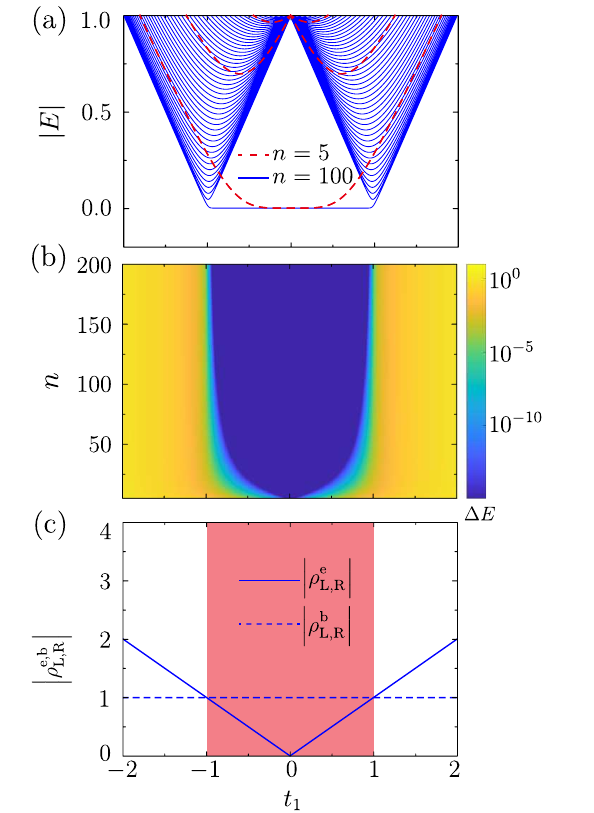}
\caption{(Color online) (a) The spectra of the Hermitian SSH model with open boundary conditions. The lengths of the chains are set as $n=5$ (the red dashed lines) and $n=100$ (the blue solid lines). (b) The finite-size energy gap $\Delta E$ in the parameter space spanned by the intra-cell hopping strength $t_1$ and chain length $n$. (c) $\left|\rho _{\text{L},\text{R}}^{\text{e},\text{b}}\right|$ as a function of $t_1$. Here, $\left|\rho _{\text{L}}^\text{e}\right|=\left|\rho _{\text{R}}^\text{e}\right|$ correspond to the blue solid line and $\left|\rho _{\text{L}}^\text{b}\right|=\left|\rho _{\text{R}}^\text{b}\right|\equiv1$ correspond to the blue dashed line. The red region determined by $\left|\rho _{\text{L},\text{R}}^{\text{e}}\right|<\left|\rho _{\text{L},\text{R}}^{\text{b}}\right|$ hosts the end states.}
\label{fig2}
\end{figure}

\subsection{Hermitian case}
\label{SHH-Hermitian}

\begin{figure}[tbp]
\includegraphics[width=8cm]{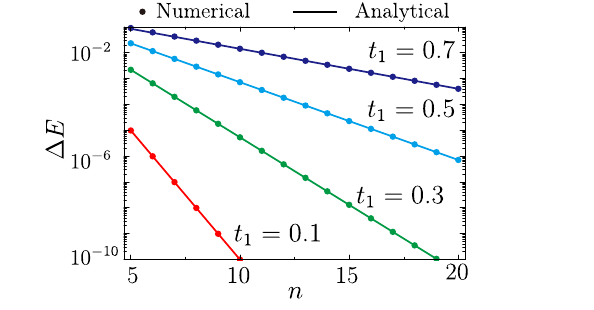}
\caption{(Color online) Semilogarithmic plot of the finite-size energy gap $\Delta E$ in the Hermitian SSH model as a function of the length of the chain for different intra-cell strengths of $t_1$. The solid dots are obtained by numerical calculations and the lines are from the analytical Eq. (\ref{SSH-gap-0}).}
\label{fig3}
\end{figure}

The Hermitian SSH model [Fig.~\ref{fig1}(a)] is perhaps the simplest model to realize topologically protected boundary states. To study the finite-size effect in the Hermitian SSH model, we set $\gamma_x=\gamma_y=\gamma_z=0$.
The localization lengths of the two end states are determined by $\left|\rho_{\text{L},\text{R}}^\text{e}\right|= t_1/t_2$. The localization lengths of the bulk states are $\left|\rho_{\text{L},\text{R}}^\text{b}\right|=1$, which implies that bulk states are the conventional Bloch waves. Figure~\ref{fig2}(c) shows the localization lengths as a function of $t_1$. From Eq.~(\ref{SSH-gap}), we get the finite-size energy gap in the Hermitian system
\begin{equation}
\Delta E=\frac{2\left(t_{2}^{2}-  t_{1}  ^{2}\right)}{t_{2}}\left(\rho _{\text{L}}^{\text{e}} \rho _{\text{R}}^{\text{e}}\right) ^{\frac{n}{2}},
\label{SSH-gap-0}
\end{equation}
We can see that the finite-size energy gap is directly related to the localization lengths of the end states.
Figure \ref{fig3} displays the finite-size energy gap obtained by analytical and numerical calculations as a function of $n$ for different $t_1$. For a given chain length, a larger $|t_1|$ will give larger $\left|\rho_{\text{L},\text{R}}^\text{e}\right|$, and thus the wave functions of the end states become more delocalized, which in turn increases the finite-size energy gap. It is found that $\Delta E$ decays exponentially with increasing $n$, and the decay rate differs for different $t_1$.

Figure~\ref{fig2}(a) illustrates the numerically calculated energy spectra as a function of the intra-cell hopping strength $t_1$ for the chain lengths $n=5$ and $n=100$. It is clear that the system can open a sizable gap for a small size $n$. In Fig.~\ref{fig2}(b), we further plot the finite-size energy gap $\Delta E$ in the space spanned by the parameters $t_1$ and $n$.

\begin{figure}[tbp]
\includegraphics[width=8cm]{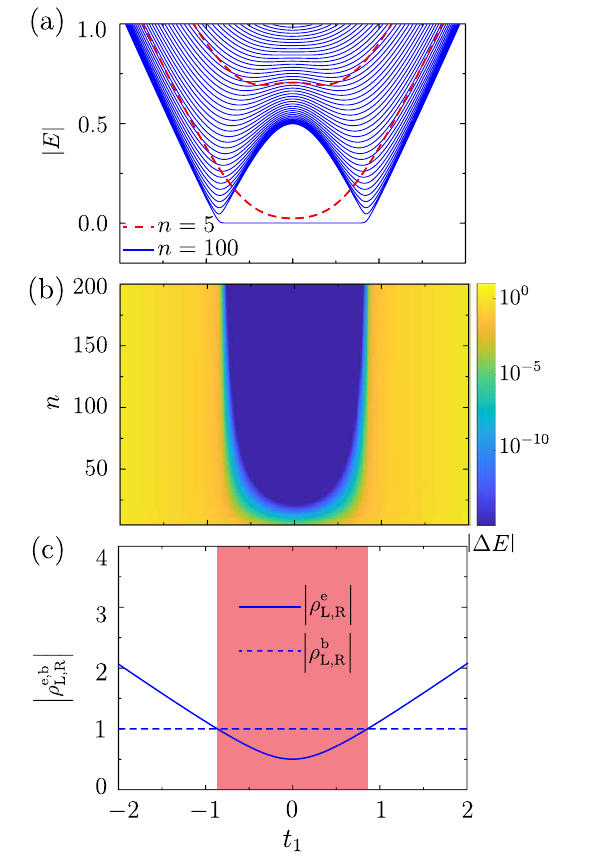}
\caption{(Color online) (a) The spectra of the non-Hermitian SSH model with an open boundary condition and $\gamma_x=0.5$, $\gamma_y=\gamma_z=0$. The lengths of the chains are set as $n=5$ (the red dashed lines) and $n=100$ (the blue solid lines). (b) The absolute value of the finite-size energy gap $\left|\Delta E\right|$ plotted in the plane formed by the intra-cell hopping strength $t_1$ and chain length $n$. (c) $\left|\rho _{\text{L},\text{R}}^{\text{e},\text{b}}\right|$  as a function of $t_1$. Here $\left|\rho _{\text{L}}^\text{e}\right|=\left|\rho _{\text{R}}^\text{e}\right|$ correspond to the blue solid line and $\left|\rho _{\text{L}}^\text{b}\right|=\left|\rho _{\text{R}}^\text{b}\right|\equiv1$ correspond to the blue dashed line. The red region is determined by $\left|\rho _{\text{L},\text{R}}^{\text{e}}\right|<\left|\rho _{\text{L},\text{R}}^{\text{b}}\right|$.}
\label{fig4}
\end{figure}

\begin{figure}[tbp]
\includegraphics[width=8cm]{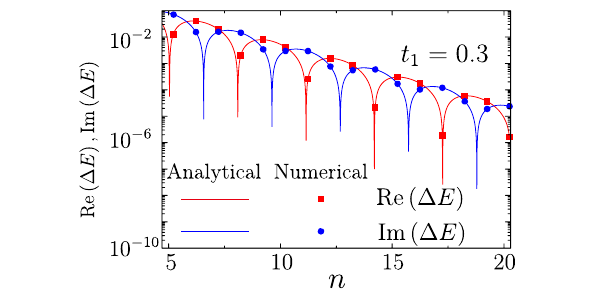}
\caption{(Color online) Semilogarithmic plot of the finite-size energy gap $\Delta E$ of the non-Hermitian SSH model with $\gamma_x=0.5$, $\gamma_y=\gamma_z=0$ for $t_1=0.3$ . The red square (blue circle) line corresponds to the real (imaginary) part of the finite-size energy gap. The dots are obtained by numerical calculations and the lines are from the analytical Eq. (\ref{SSG-gap-x}).}
\label{fig5}
\end{figure}

\subsection{Non-Hermitian case of $\gamma_x\neq 0,\gamma_y=\gamma_z=0$}
\label{SHH-gammax}

Now we investigate the finite-size  effect in the non-Hermitian SSH system with only the non-conjuated intra-cell hopping $\gamma_{x}$ [Fig.~\ref{fig1}(b)]. In this case, $\rho _{\text{L},\text{R}}^{\text{e}}= -\left(t_1+i\gamma_x\right)/{t_{2}}$, and thus, the localization lengths of end states determined by $\left|\rho_{\text{L},\text{R}}^{\text{e}}\right|=\sqrt{t_1^2+\gamma_x^2}/t_{2}$ increase due to the existence of $\gamma_{x}$, while $\left|\rho_{\text{L},\text{R}}^\text{b}\right|$ keep unchanged. Because of this, the region for end states becomes narrow in the presence of $\gamma_{x}$ as shown in Fig.~\ref{fig4}(c).

The interference pattern of the end-state wave function in Eq.~(\ref{SSH-wave}) varies, and the finite-size energy gap $\Delta E$ turns to be
\begin{equation}
\Delta E=\frac{2\left[t_{2}^{2}-\left(  t_{1}+i\gamma_{x}\right)  ^{2}\right]}{t_{2}}\left(\rho _{\text{L}}^{\text{e}} \rho _{\text{R}}^{\text{e}}\right) ^{\frac{n}{2}}.
\label{SSG-gap-x}
\end{equation}%
The finite-size energy gap becomes complex due to $\gamma_{x}$. As shown in Fig.~\ref{fig5}, both the real and imaginary parts of $\Delta E$ exhibit an oscillating exponential decay with increasing the chain length, and the oscillation period is $T=2\pi /\arg{\left(\rho_\text{L}^\text{e} \rho_\text{R}^\text{e}\right)}$. This is distinct from the Hermitian case where the finite-size energy gap is real and shows a monotonic-exponential decay as the chain length increases. The consistency of the numerical and analytical calculations demonstrates reliability of the obtained results.

In Fig.~\ref{fig4}(a), we sketch the absolute value of the numerical eigenenergies as a function of the intra-cell hopping strength $t_1$ for the chain lengths $n=5$ and $n=100$ when $\gamma_x= 0.5$, $\gamma_y=\gamma_z=0$. The absolute value of the finite-size energy gap $\left|\Delta E\right|$ is plotted in the parameter space formed by $t_1$ and $n$ as shown in Fig.~\ref{fig4}(b). We can see that the results are found quite similar to those in the Hermitian case in Sec.~\ref{SHH-Hermitian} except that the region for end states is narrower and the finite-size energy gap can even appear at $t_1=0$.

At last, we would like to point out that, although there exits the non-Hermitian term $\gamma_{x}$, the bulk states are still captured by the conventional Bloch waves since $\left|\rho_{\text{L},\text{R}}^\text{b}\right|= 1$, which implies the non-Hermitian skin effect \cite{Yao2018PRLa} is absent in this case.

\subsection{Non-Hermitian case of $\gamma_y\neq 0,\gamma_x=\gamma_z=0$}
\label{SHH-gammay}
\begin{figure}[tbp]
\includegraphics[width=8cm]{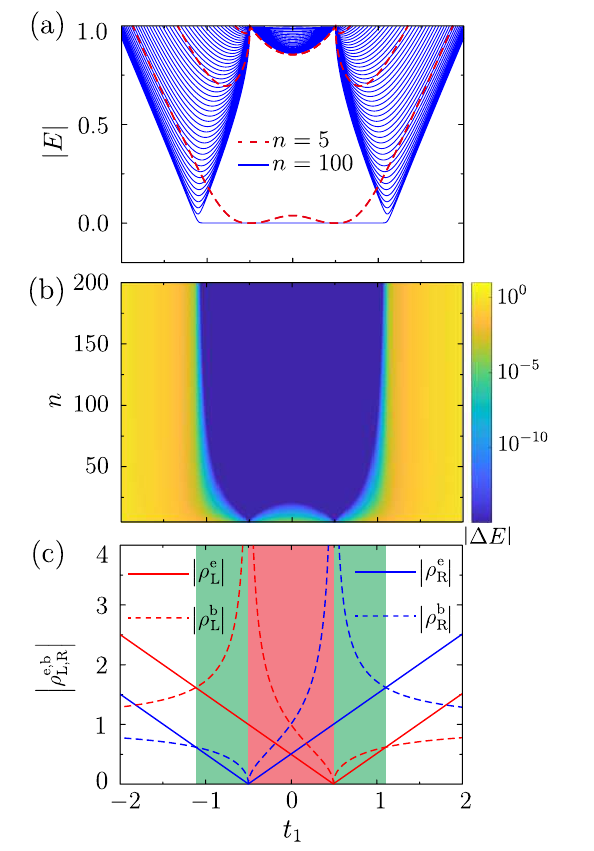}
\caption{(Color online) (a) The spectra of the non-Hermitian SSH model with an open boundary condition and $\gamma_y=0.5$, $\gamma_x=\gamma_z=0$. The lengths of the chains are set as $n=5$ (the red dashed lines) and $n=100$ (the blue solid lines). (b) The absolute value of the finite-size energy gap $\left|\Delta E\right|$ plotted in the parameter space formed by the intra-cell hopping strength $t_1$ and chain length $n$. (c) $\left|\rho _{\text{L},\text{R}}^{\text{e},\text{b}}\right|$ as a function of $t_1$. Here, $\left|\rho _{\text{L},\text{R}}^\text{e}\right|$ correspond to the red and blue solid lines and $\left|\rho _{\text{L},\text{R}}^\text{b}\right|$ correspond to the red and blue dashed lines. The red region is given by $\left|\rho _{\text{L},\text{R}}^{\text{e}}\right|<1<\left|\rho _{\text{L},\text{R}}^{\text{b}}\right|$, while the green region are determined by $1<\left|\rho_{\text{L}}^{\text{e}}\right|<\left|
\rho_{\text{L}}^{\text{b}}\right|$ or $1<\left|\rho_{\text{R}}^{\text{e}}\right|<
\left|\rho_{\text{R}}^{\text{b}}\right|$}
\label{fig6}
\end{figure}

\begin{figure}[tbp]
\includegraphics[width=8cm]{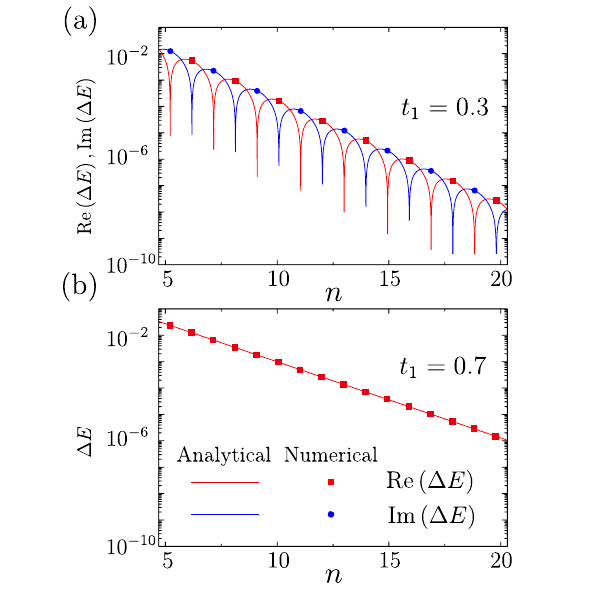}
\caption{(Color online) Semilogarithmic plot of the finite-size energy gap $\Delta E$ of the non-Hermitian SSH model with $\gamma_y=0.5$, $\gamma_x=\gamma_z=0$ for (a) $t_1=0.3$ and (b) $t_1=0.7$. The red square (blue circle) line corresponds to the real (imaginary) part of the finite-size energy gap. The dots are obtained by numerical calculations and the lines are from by analytical Eq.~(\ref{SSG-gap-y}). Note that the imaginary part of $\Delta E$ in (b) is vanishing, so we only show the real part of $\Delta E$.}
\label{fig7}
\end{figure}

Here we investigate the finite-size effect in the SSH model with unequal intra-cell hoppings $\gamma_y$ [see Fig.~\ref{fig1}(c)].
We found that $\gamma_y$ modifies the end-state localization lengths to be $ \left| \rho_{\text{L},\text{R}}^{\text{e}}\right|=\left|\left(-t_{1}\pm \gamma_{y}\right)/{t_{2}} \right|$,
which indicates that the left and right end states are most localized at $t_1=\gamma_y$ and $t_1=-\gamma_y$, respectively. Interestingly, $\gamma_{y}$ causes an asymmetry in $\rho_{\text{L},\text{R}}$ for both the end and bulk states [see Fig.~\ref{fig6}(c)]. For the bulk states, $\left|\rho_{\text{L}}^{\text{b}}\right|=1/\left|\rho_{\text{R}}^\text{b}\right|=\left|\sqrt{\left(t_{1}-\gamma_{y}\right)/\left(t_{1}+\gamma_{y}\right)}\right|$. 
When $t_1>0$, $\left|\rho_{\text{L}}^{\text{b}}\right|<1$ and $\left|\rho_{\text{R}}^{\text{b}}\right|>1$, this indicates that the bulk states are localized at the left boundary. When $t_1<0$, the bulk states are localized at the right boundary as $\left|\rho_{\text{L}}^{\text{b}}\right|>1$ and $\left|\rho_{\text{R}}^{\text{b}}\right|<1$. This is called the non-Hermitian skin effect~\cite{Yao2018PRLa}. The two end states are separately located at the two opposite sides when $\left|\rho_{\text{L},\text{R}}^{\text{e}}\right|<1<\left|\rho_{\text{L},\text{R}}^{\text{b}}\right|$, both located at the left side when $1<\left|\rho_{\text{L}}^{\text{e}}\right|<\left|\rho_{\text{L}}^{\text{b}}\right|$, and both located at the right side when $1<\left|\rho_{\text{R}}^{\text{e}}\right|<\left|\rho_{\text{R}}^{\text{b}}\right|$.\cite{Yao2018PRLa}

The finite-size  energy gap is given by
\begin{equation}
\Delta E=\frac{2\left[t_{2}^{2}-\left(
	t_{1}^{2}-\gamma_{y}^{2}\right) \right] }{t_{2}}\left(\rho _{\text{L}}^{\text{e}} \rho _{\text{R}}^{\text{e}}\right) ^{\frac{n}{2}},%
\label{SSG-gap-y}
\end{equation}
which implies an oscillating exponential decay when $\left|t_1\right|<\left|\gamma_y\right|$ with the period $T=2\pi / \arg{\left(\rho_\text{L}^\text{e} \rho_\text{R}^\text{e}\right)=2}$, and a monotonic exponential decay when $\left|t_1\right|>\left|\gamma_y\right|$.

In Fig.~\ref{fig6}(a), we plot the absolute value of numerical eigenvalues as a function of $t_1$ for $n=5$ and $n=100$. The absolute value of the finite-size energy gap $\left|\Delta E\right|$ in the parameter space spanned by $t_1$ and $n$ is shown in Fig.~\ref{fig6}(b). When the chain is short, $\left|\Delta E\right|$ shows a local maximum value at $t_1=0$. This is different from the Hermitian case and the non-Hermitian case with only $\gamma_x$ term, in which $\left|\Delta E\right|$ is smallest at $t_1=0$ for a short chain. The reason is that, due to $\gamma_{y}$, the minimum of the localization lengths of the end states deviates from $t_1=0$ and splits into two minima as shown in Fig.~\ref{fig6}(c). Figure~\ref{fig7} displays the finite-size  energy gap as a function of the chain length $n$. Upon increasing  $n$, both the real and imaginary parts of the finite-size energy gap $\Delta E$ show an oscillating exponential decay when $\gamma_{y}>t_1=0.3$ [see Fig.~\ref{fig7}(a)] and $\Delta E$ can be fully real or imaginary depending on whether $n$ is even or odd number. When $t_1=0.7>\gamma_{y}$, the gap is fully real, and the oscillation disappears [see Fig.~\ref{fig7}(b)].

\subsection{Non-Hermitian case of $\gamma_z\neq 0,\gamma_x=\gamma_y=0$}
\label{SHH-gammaz}
\begin{figure}[tbp]
\includegraphics[width=8cm]{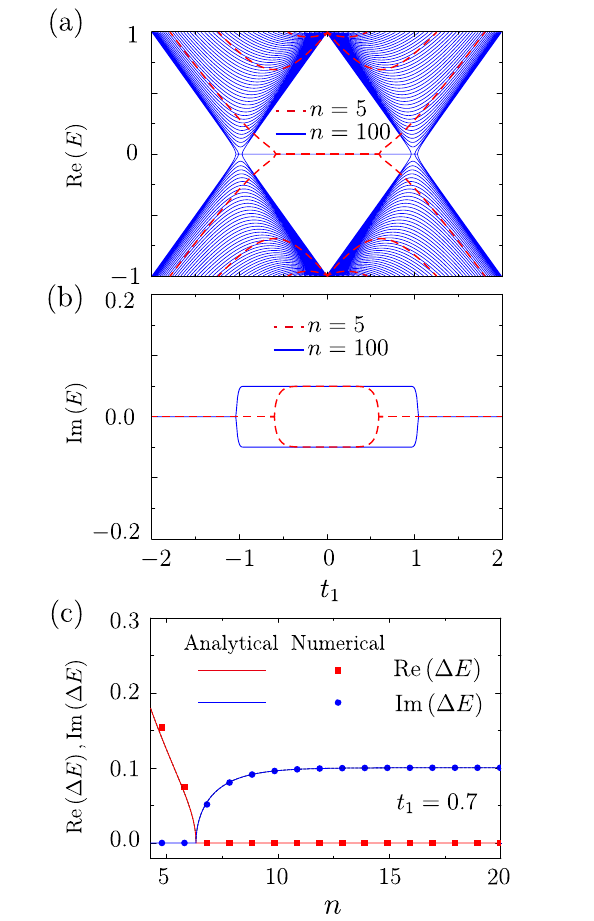}
\caption{(Color online) (a) and (b) corresponding to the real and imaginary parts of the spectra in the non-Hermitian SSH model with an open boundary condition and $\gamma_z=0.05$, $\gamma_x=\gamma_y=0$. The lengths of the chains are set as $n=5$ (the red lines) and $n=100$ (the blue lines). In (b) we only show the imaginary part of the spectra of the end states. (c) The real (red square line) and the imaginary (blue circle line) of the finite-size energy gap $\Delta E$ for $t_1=0.7$. The dots are obtained by numerical calculations and the lines are from by analytical Eq. (\ref{SSG-gap-z}). }
\label{fig8}
\end{figure}

In this subsection, we consider the SSH model with an imaginary staggered on-site potential $\gamma_{z}$. As pointed out in Section \ref{SSH-Model}, $\gamma_{z}$ breaks chiral symmetry but not $PT$ symmetry. From the analytical results, we find that the $\gamma_z$ has no influence on $\rho_{\text{L},\text{R}}^{\text{e},\text{b}}$, which means that the localization lengths of the system remain unchanged. However, the finite-size energy gap becomes
\begin{equation}
\Delta E=2\sqrt{\frac{\left(t_{2}^{2}-t_{1}%
		^{2}\right)^2}{t_{2}^2}\left(\rho _{\text{L}}^{\text{e}} \rho _{\text{R}}^{\text{e}}\right) ^{n}-\gamma_{z}^{2}}.
\label{SSG-gap-z}
\end{equation}

When $n$ is small, for $t_s<\left|t_1\right|<t_2$ ($t_s$ determined by a transcendental equation $\left(t_{2}^{2}-t_{s}^{2}\right)^2(t_s/t_2)^{2n}=t_2^2\gamma_z^2$), the number within the square root of Eq.~(\ref{SSG-gap-z}) is positive, and therefore $\Delta E$ is purely real. However, for $\left|t_1\right|<t_s<t_2$, the number within the square root becomes negative, so $\Delta E$ is purely imaginary. When $n$ is large enough, for finite $\gamma_z$, the number within the square root of Eq.~(\ref{SSG-gap-z}) is negative since $\left(\rho _{\text{L}}^{\text{e}} \rho _{\text{R}}^{\text{e}}\right) ^{n}$ tends to be zero, and thus $\Delta E$ becomes a purely imaginary number $2i\gamma_z$ which is independent of $t_1$.

In Figs.~\ref{fig8}(a-b), we separately sketch the real and imaginary parts of the energy spectra as a function of $t_1$ for $n=5$ and $n=100$. These results are in good consistent with the previous analysis based on Eq.~(\ref{SSG-gap-z}). Figure~\ref{fig8}(c) illustrates the real and imaginary parts of the finite-size energy gap obtained by the analytical and numerical calculations as a function of the chain length $n$ for $t_1=0.7$. We can see that the finite-size energy gap goes though a transition from real values to imaginary values.

So far, we have plotted the real and imaginary parts of the complex finite-size energy gap as a function of the chain length separately. To look at the finite-size energy gap from a different point of view, we also present the energy splitting due to the hybridization of end states on the complex energy plane in Appendix \ref{Appendix_complex_plane}.

\section{Non-Hermitian Chern insulator model}
\label{QWZ}
In this section, we study the finite-size  effect in 2D non-Hermitian topological systems. As a concrete example, we consider the following Chern insulator model Hamiltonian \cite{Qi2006PRB}
\begin{equation}
H_{\text{CI}}\left( k_x,k_y\right) =A\left(k_x\sigma _{x}+k_y\sigma _{y}\right)+(M-Bk^2)\sigma _{z},
\end{equation}%
where $k^2=k_x^2+k_y^2$, $A,M,B$ are model parameters, $\sigma_{x,y,z}$ are the Pauli matrices acting on the spin or orbital space. The non-Hermitian Hamiltonian is expressed as  $H_{\gamma}=i\left( \gamma _{x}\sigma _{x}+\gamma _{y}\sigma _{y}+\gamma_{z}\sigma _{z}\right)$. Then the total Hamiltonian of the non-Hermitian Chern insulator model is $H=H_{\text{CI}}+H_\gamma$.

To get the information about the localization length and energy spectrum, following from Ref. \cite{Zhou2008PRL}, we solve the non-Hermitian Chern insulator model $H$ in a finite strip geometry of the width $L$ with a periodic boundary condition along the $x$ direction and an open boundary condition along the $y$ direction. $k_x$ is still a good quantum number, while $k_y$ is replaced by using the Peierls substitution, $k_y=-i\partial_y$. In the analytical calculations, we ignore $\gamma_y$ for simplicity. We can rewrite $H$ as
\begin{equation}
H_{1}\left( k_x,-\partial_y\right) =A\left(\tilde{k}_x\sigma _{x}-\partial_y\sigma _{y}\right)+\left[\tilde{M}-B\left(\tilde{k}_x^2-\partial_y^2\right)\right]\sigma _{z},
\end{equation}%
where $\tilde{k}_x=k_x+i \gamma_x/A$ and $\tilde{M}\left(k_x\right)=M+i \gamma_z-B\gamma_x\left(\gamma_x-2iAk_x\right)/A^2$.

For a strip geometry with the boundary conditions $\Phi\left(k_x,y=\pm L/2\right)=0$, the spectrum is given by solving the following equation\cite{Zhou2008PRL}:
\begin{equation}
\frac{\tanh\frac{\lambda_{1}L}{2}}{\tanh\frac{\lambda_{2}L}{2}}+\frac
{\tanh\frac{\lambda_{2}L}{2}}{\tanh\frac{\lambda_{1}L}{2}}=\frac{\alpha
_{1}^{2}\lambda_{2}^{2}+\alpha_{2}^{2}\lambda_{1}^{2}-k_{x}^{2}\left(
\alpha_{1}-\alpha_{2}\right)  ^{2}}{\alpha_{1}\alpha_{2}\lambda_{1}\lambda
_{2}},%
\label{gapQWZ}
\end{equation}
where $\alpha_{1,2}=E-\tilde{M}+B\left(  \tilde{k}_{x}^{2}-\lambda_{1,2}^{2}\right), \lambda_{1,2}^{2}=\tilde{k}_{x}^{2}+F\pm\sqrt{F^{2}-\left(  \tilde{M}^{2}%
-E^{2}\right)  /B^{2}}$, and $F=\left(  A^{2}-2\tilde{M}B\right)  /\left(
2B^{2}\right) $. The corresponding wave function distribution has the form $\Phi=\sum_{\eta,i}C_{\eta,i}\left(
\begin{array}{c}
\mu _{i} \\
\upsilon _{i}%
\end{array}%
\right) e^{\eta\lambda_{i}}$, where $\eta=\pm$, $i=1,2$, $\mu _{i}=\tilde{M}-B\left( \tilde{k}_{x}^{2}-\lambda
_{i}^{2}\right) +E$, $\upsilon _{i}=A\left( \tilde{k}_{x}-\lambda _{i}\right)$, and $C_{\eta,i}$ are normalization constants.
Here $\left|\lambda_{1,2}\right|$ determine the localization lengths of the edge states in the Chern insulator model, which play the same roles as $\left|\rho_{\text{L},\text{R}}^{\text{e}}\right|$ in the 1D SSH model. In the large $L$ limit near $k_x=0$, we have
\begin{equation}
E_{\pm}=\pm A\tilde{k}_x.
\label{QWZsemi}
\end{equation}%
Therefore the edge-state energy can be modified by $\gamma_x$ and $\lambda_{1,2}$ can be affected by both $\gamma_x$ and $\gamma_z$.

Since the finite-size effect in the Hermitian Chern insulator model has been well studied in the literature\cite{Zhou2008PRL}, we just review the main results. By expanding Eq.~(\ref{gapQWZ}) at $k_x=0$ for $E=0$ and assuming $\operatorname{Re}\left(\lambda_1\right)\gg\operatorname{Re}\left(\lambda_2\right)$, the finite-size energy gap as a function of the strip width $L$ is given as\cite{Zhou2008PRL,Shan2010NJP}
\begin{equation}
\Delta E=\left\{
\begin{array}{l}
\frac{4AM}{\sqrt{A^{2}-4BM}}e^{-\lambda _{2}L} \\
\frac{4AM}{\sqrt{4BM-A^{2}}}e^{-\operatorname{Re}\left(\lambda_2\right) L}\sin \left( \operatorname{Im}\left(\lambda_2\right)L\right)
\end{array}%
\left.
\begin{array}{c}
A^{2}>4BM, \\
A^{2}<4BM.%
\end{array}%
\right. \right.
\label{QWZ-hermitian}
\end{equation}%
 For $A^2>4BM$, $\lambda_2$ is a real number, and the finite-size energy gap $\Delta E$ is found to exhibit a monotonic exponential decay with increasing system size, while for the case of $A^2<4BM$, $\lambda_2=A/2B+i\sqrt{4BM-A^{2}}/2B$ becomes a complex number, and $\Delta E$ exhibits an oscillating exponential decay.

 In the following calculations, we will consider the case of $A^2>4BM$ and choose $A=B=1$ and $M=0.1$. For the non-Hermitian Chern insulator model, it is also found that the gap always open at $k_x=0$. Therefore, we will concentrate on the point $k_x=0$.

 It is worth noting that, for $k_x=0$, the non-Hermitian Chern insulator model becomes another version of non-Hermitian SSH model.

\subsection{Non-Hermitian case of $\gamma_x\neq0,\gamma_y=\gamma_z=0$}
\begin{figure}[tbp]
\includegraphics[width=8cm]{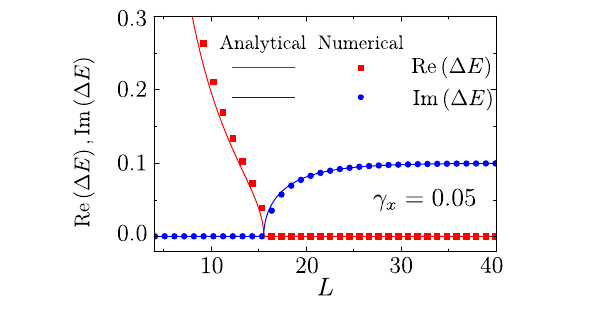}
\caption{(Color online) The real part (red square line) and the imaginary part (blue circle line) of the finite-size energy gap $\Delta E$ in the non-Hermitian Chern insulator model for $\gamma_x=0.05,\gamma_y=\gamma_z=0$. The dots are obtained by numerical calculations and the lines are from by analytical Eq. (\ref{gapQWZ}).}
\label{fig9}
\end{figure}
Here we investigate the finite-size effect in the non-Hermitian Chern insulator model with only $\gamma_x$ term, and set $\gamma_y=\gamma_z=0$. Figure~\ref{fig9} illustrates the real and imaginary parts of the finite-size energy gap $\Delta E$ as a function of the strip width $L$ for $\gamma_x=0.05$. Upon increasing $L$, the finite-size energy gap decreases from a purely real number to zero, at a certain $L$ the imaginary part is developed and reaches a constant value $2\gamma_x$ as indicated by Eq. (\ref{QWZsemi}). The result is quite similar to that in the non-Hermitian SSH model with only $\gamma_z$ [Sec.~\ref{SHH-gammaz}] since in this case, for $k_x=0$, the non-Hermitian Chern insulator model is equivalent to the non-Hermtian SSH model with $PT$ symmetry.

\subsection{Non-Hermitian case of $\gamma_y\neq0,\gamma_x=\gamma_z=0$}
\begin{figure}[tbp]
\includegraphics[width=8cm]{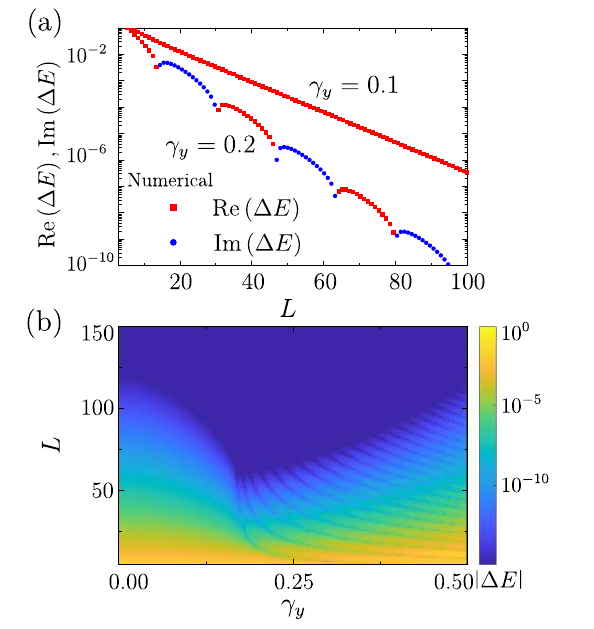}
\caption{(Color online) (a) Semilogarithmic plot of the finite-size energy gap $\Delta E$ in the non-Hermitian Chern insulator model as a function of the strip width $L$ for $\gamma_y=0.05,\gamma_x=\gamma_z=0$. The red square (blue circle) dots correspond to the real (imaginary) part of the finite-size energy gap. (b) The absolute value of the finite-size energy gap $\Delta E$ in the plane formed by $\gamma_y$ and $L$. }
\label{fig10}
\end{figure}

Now we study the non-Hermitian case of $\gamma_y$, and set $\gamma_x=\gamma_z=0$. In Fig.~\ref{fig10}(a), we plot the real and imaginary parts of the finite-size energy gap $\Delta E$ as a function of the strip width $L$. For $\gamma_y=0.1$, the edge states are purely real, and $\Delta E$ decays exponentially with $L$. However, for a larger $\gamma_y=0.2$, upon increasing strip width, $\Delta E$ shows an oscillating exponential decay and alternates between real and imaginary values. $\left|\Delta E\right|$ in the $\gamma_y$-$L$ parameter space obtained by numerical simulations are shown in Fig.~\ref{fig10}(b), from which we can clearly see the oscillatory pattern of $\left|\Delta E\right|$ for lager $\gamma_y$. The results are also similar to to those in the non-Hermitian SSH model with $\gamma_y$ [Sec.~\ref{SHH-gammay}] because in this case, for $k_x=0$, the non-Hermitian Chern insulator model is equivalent to the non-Hermitian SSH model with chiral symmetry.

\subsection{Non-Hermitian case of $\gamma_z\neq0,\gamma_x=\gamma_y=0$}
\begin{figure}[tbp]
\includegraphics[width=8cm]{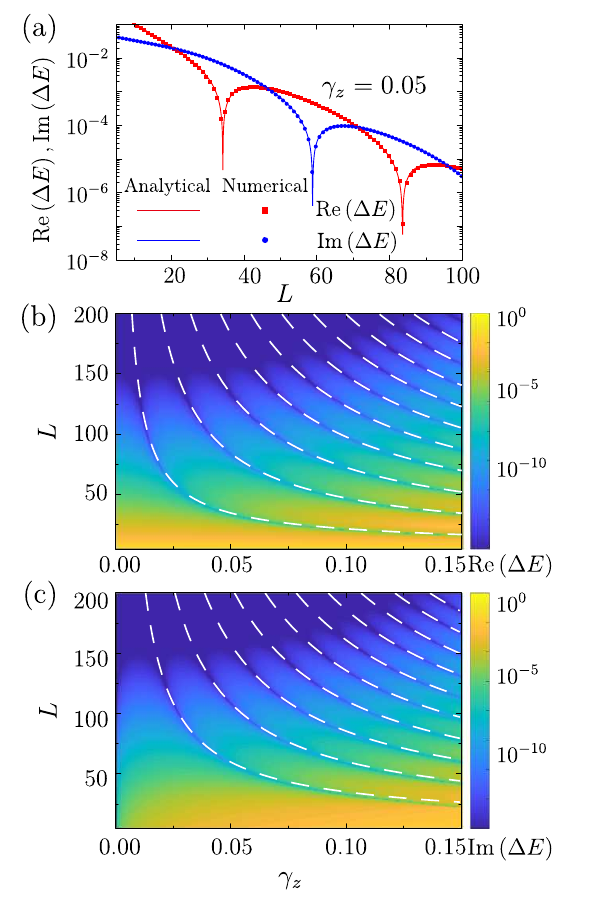}
\caption{(Color online) (a) Semilogarithmic plot of the finite-size energy gap $\Delta E$ for the non-Hermitian Chern insulator model with $\gamma_z=0.05,\gamma_x=\gamma_y=0$. The red square (blue circle) line corresponds to the real (imaginary) part of the finite-size energy gap. The dots are obtained by numerical calculations and the lines are from analytical Eq. (\ref{gapQWZ}). (b) and (c) are the real and imaginary parts of the energy gap $\Delta E$ in the parameter space of $\gamma_z$ and $n$. The white dashed lines are obtained by analytical Eq. (\ref{dipposition}), and they depict the positions of the dips in the finite-size  energy gap. }
\label{fig11}
\end{figure}

In the last subsection, we consider the non-Hermitian Chern insulator model with only $\gamma_z$. Figure~\ref{fig11}(a) shows the real and imaginary parts of the finite-size energy gap varying as a function of the strip width $L$ for $\gamma_z=0.05$. We can see that both the real and imaginary parts show an oscillating exponential decay. The real and imaginary parts of the gap $\operatorname{Re}\left(\Delta E\right)$ and $\operatorname{Im}\left(\Delta E\right)$ in the $\gamma_y$-$L$ parameter space are shown in Figs.~\ref{fig11}(b-c). 

Using the same method as in Eq. \eqref{QWZ-hermitian}, we obtain
\begin{equation}
\Delta E   =\frac{4A\left(  M+i\gamma_{z}\right)  }{\sqrt{A^{2}-4B\left(  M+i\gamma
_{z}\right)  }}e^{-\lambda_2 L},
\end{equation}
where $\lambda_2=\left(  2A-\sqrt{2}T\right)  /4B+i\sqrt{2}\gamma_{z}/T$ and $T=\sqrt
{A^{4}-4BM+\sqrt{\left(  A^{2}-4BM\right)  ^{2}+16B^{2}\gamma_{z}^{2}}}$. The gap is found to be a complex value, and its real and imaginary parts are given by
\begin{align}
\operatorname{Re}\left(\Delta E\right)   & =\frac{4A\left(  M\cos\theta+\gamma_{z}\sin
\theta\right)  }{\left[  \left(  A^{2}-4BM\right)  ^{2}+16B^{2}\gamma_{z}%
^{2}\right]  ^{1/4}},\nonumber \\
\operatorname{Im}\left(\Delta E\right)  & =\frac{4A\left(  M\sin\theta-\gamma_{z}\cos
\theta\right)  }{\left[  \left(  A^{2}-4BM\right)  ^{2}+16B^{2}\gamma_{z}%
^{2}\right]  ^{1/4}},
\label{dipposition}
\end{align}
with $\theta=\operatorname{Im} \left(\lambda_2\right)L+\theta_{0}$, and $\theta_{0}=\arg\left[  A^{2}-4B\left(  M+i\gamma
_{z}\right)  \right]  $. By calculating $\operatorname{Re}\Delta E=0$ and $\operatorname{Im}\Delta E=0$, we obtain
\begin{align}
L_{\operatorname{Re}}  & =\frac{N\pi-\arctan\left(  M/\gamma_{z}\right)
-\theta_{0}}{\operatorname{Im} \left(\lambda_2\right)},\nonumber \\
L_{\operatorname{Im}}  & =\frac{N\pi+\arctan\left(  M/\gamma_{z}\right)
-\theta_{0}}{\operatorname{Im} \left(\lambda_2\right)},
\end{align}
where $N=1,2,\cdots$, and $L_{\operatorname{Re,Im}}$ correspond to the widths where the real and imaginary parts of the edge-state energy are zero. The oscillation period is $T=1/{\operatorname{Im} \left(\lambda_2\right)}$. Figures~\ref{fig11}(b-c) demonstrate that the numerical results are in good agreement with the analytical results.

Therefore we conclude that the non-Hermitian term $\gamma_z$ introduces an imaginary part to $\lambda_2$, adjusting the localization lengths of the system. The value of the finite-size energy gap becomes complex and exhibits an oscillating exponential decay with increasing $L$ [Fig.~\ref{fig11}]. This result is also similar to that in the chiral SSH model with only $\gamma_x$  [Sec.~\ref{SHH-gammax}].

\section{Conclusion}
\label{Conclusion}

To conclude, we investigated the finite-size effect in the non-Hermitian 1D SSH model. We have shown that the non-Hermitian hopping terms $\gamma_{x}$ and $\gamma_{y}$ that respect chiral symmetry can modify the end-state localization and cause a complex finite-size energy gap that exhibits an oscillating exponential decay with increasing the chain length. However, the imaginary staggered on-site potential that breaks chiral symmetry can produce a finite-size energy gap transition from real values to imaginary values. In addition, we also studied the finite-size  effect in a 2D non-Hermitian Chern insulator model and found the similar behaviors as those in the 1D non-Hermitian SSH model.

\section*{Acknowledgments}
We would like to thank Hui-Ke Jin for helpful discussions. R.C. and D.-H.X. were supported by the National Natural Science Foundation of China (Grant No. 11704106) and the Scientific Research Project of Education
Department of Hubei Province (Grant No. Q20171005). D.-H.X. also acknowledges the support of the
Chutian Scholars Program in Hubei Province.

\appendix
\section{Recursion relation of the secular equation in the presence of next nearest neighbor hoppings}
\label{Appendix_NNN}

In Sec.~\ref{SSH-Model}, we have given the recursion relation of the secular equation for a non-Hermitian SSH model with nearest neighbor hoppings. Here we present the recursion relation for a finite non-Hermitian SSH chain including next nearest neighbor hoppings. The corresponding Schr\"{o}dinger equation is
\begin{equation}
\left( H_{n}^{F}-E \right) \Psi ^n= \begin{pmatrix}
	\epsilon&		T&		&		&		\\
	T^{\dag}&		\epsilon&		T&		&		\\
	&		\ddots&		\ddots&		\ddots&		\\
	&		&		T^{\dag}&		\epsilon&		T\\
	&		&		&		T^{\dag}&		\epsilon\\
\end{pmatrix}
\begin{pmatrix}
	\Psi _{1}^{n}\\
	\Psi _{2}^{n}\\
	\ddots\\
	\Psi _{n-1}^{n}\\
	\Psi _{n}^{n}\\
\end{pmatrix}  =0,
\end{equation}
where $
\epsilon =\begin{pmatrix}
	-E&		t_+\\
	t_-&		-E\\
\end{pmatrix}
$, $
T=\begin{pmatrix}
	0&		t_3\\
	t_2&		0\\
\end{pmatrix}
$, $
t_{\pm}=t_1+i\gamma _x\pm \gamma _y
$ and $t_3$ represents the next nearest neighbor hopping. Denoting the determinant of the secular equation as $F\left( n \right) =\left| H_{n}^{\text{F}}-E \right|$, and $F$ satisfies the following recursion relation
\begin{equation}
\mathcal{Z}\left( i \right)=
\begin{pmatrix}
	F\left( i \right)\\
	G\left( i \right)\\
	H\left( i \right)\\
	I\left( i \right)\\
	J\left( i \right)\\
	K\left( i \right)\\
\end{pmatrix} = T  \begin{pmatrix}
	F\left( i-1 \right)\\
	G\left( i-1 \right)\\
	H\left( i-1 \right)\\
	I\left( i-1 \right)\\
	J\left( i-1 \right)\\
	K\left( i-1 \right)\\
\end{pmatrix}=T\mathcal{Z}\left( i-1 \right)  ,
\end{equation}
where $G\left(i\right),\cdots,K\left(i\right)$ are intermediate variables, and
\begin{equation}
T=
\begin{pmatrix}
	E^2-t_+t_-&		Et_2&		-t_3t_-&		t_2t_+&		-Et_3&		-t_2t_3\\
	-Et_2&		-t_{2}^{2}&		0&		0&		0&		0\\
	t_2t_-&		0&		0&		-t_{2}^{2}&		0&		0\\
	-t_3t_+&		0&		-t_{3}^{2}&		0&		0&		0\\
	Et_3&		0&		0&		0&		-t_{3}^{2}&		0\\
	-t_2t_3&		0&		0&		0&		0&		0\\
\end{pmatrix}.
\end{equation}
The recursion relation can also expressed as
\begin{align}
\mathcal{Z}\left( i \right) =Q\Lambda Q^{-1}\mathcal{Z}\left( i-1 \right),
\end{align}
where we have used the eigen-decomposition of the matrix $T=Q\Lambda Q^{-1}$, $Q$ is a square $N\times N$ matrix whose $i$-th column is the eigenvector $q_i$ of $T$, and $\Lambda$ is a diagonal matrix whose diagonal elements are the corresponding eigenvalues.

\section{The spectra of end states on the complex energy plane}
\label{Appendix_complex_plane}

\begin{figure}[b]
\includegraphics[width=8cm]{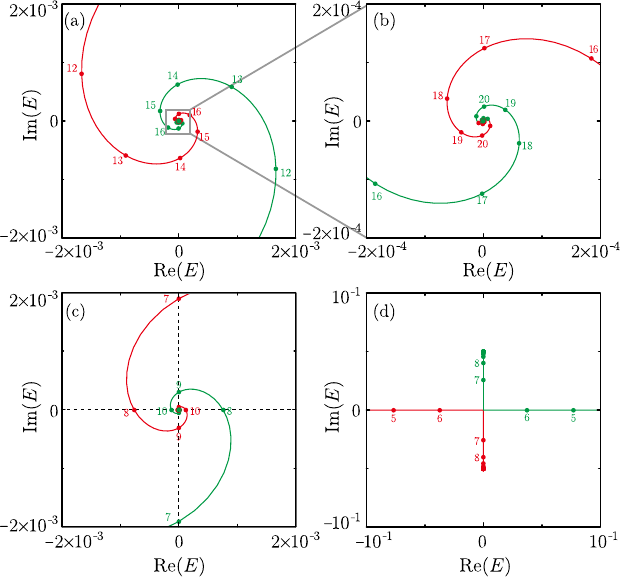}
\caption{Splitting energies (the red and green dotted lines) on the complex energy plane due to the hybridization of the two end states as functions of the chain length $n$. The dots are obtained by numerical calculations and the lines are from analytical expression Eq. (\ref{SSH-gap}). The two spiral lines formed by splitting energies eventually converge at the origin of the complex plane, which implies the finite-size gap is closed for large system size.  Here (a) corresponds to Fig.~\ref{fig5} with $\gamma_x=0.5$, $\gamma_y=\gamma_z=0$, $t_1=0.3$, (b) is a zoom-in version of (a), (c) corresponds to Fig.~\ref{fig7}(a) with $\gamma_y=0.5$, $\gamma_x=\gamma_z=0$, $t_1=0.3$, and (d) corresponds to Fig.~\ref{fig8}(c) with $\gamma_z=0.05$, $\gamma_x=\gamma_y=0$, $t_1=0.7$. }
\label{fig12}
\end{figure}

In Sec.~\ref{SSH-Results}, we have separately plotted the real and imaginary parts of the finite-size gap as functions of the chain length $n$ for the non-Hermitian SSH model. While the spectra on the complex energy plane may provide a different angle on non-Hermitian properties of topological systems~\cite{Lee2018arXiv1,Gong2018PRX,Yao2018PRLb,Shen2018PRL}. As shown in Fig.~\ref{fig12}, we plot the splitting energies due to the hybridization of end states as functions of the chain length $n$ on the complex energy plane.

In the case of $\gamma_x\neq 0,\gamma_y=\gamma_z=0$~[Fig.~\ref{fig12}(a-b)], the splitting energy evolution as a function of the chain length shows spiral lines on the complex plane. This is different from the Hermitian case where the splitting energies are only located on the real axes. In the case of $\gamma_y\neq 0,\gamma_x=\gamma_z=0$ with $\left|t_1\right|<\left|\gamma_y\right|$~[Fig.~\ref{fig12}(c)], the results are similar to the former case, except that the splitting energies alternate between real and imaginary values for discrete chain length $n$. In the case of $\gamma_z\neq 0,\gamma_x=\gamma_y=0$~[Fig.~\ref{fig12}(d)], the splitting energies are purely real when $n$ is small, and become purely imaginary for large $n$.

In addition, for the non-Hermitian Chern insulator model, the splitting energies (at $k_x=0$) on the complex plane demonstrate similar behavior.

\bibliography{bibfile}

\end{document}